\documentclass[aps,pre,twocolumn,amsmath,amssymb,nofootinbib,superscriptaddress,showpacs,longbibliography]{revtex4-2}
\pdfoutput=1
\usepackage[utf8]{inputenc}
\usepackage[english]{babel}
\usepackage[T1]{fontenc}
\usepackage{amsmath}
\usepackage{hyperref}

\newcommand{\tr}[1]{\mathrm{tr}(#1)}

\newcommand{\Var}[1]{\mathrm{var}\left[#1\right]}
\newcommand{\Cov}[1]{\mathrm{cov}\left(#1\right)}

\usepackage{braket}
\usepackage{amsmath, amssymb, amsthm}
\usepackage{natbib}
\usepackage{mathtools}
\usepackage{xcolor}
\usepackage{comment}
\usepackage{wasysym}
\usepackage{slashed}

\usepackage{tikz}
\usepackage{lipsum}

\begin{document}
	
\title{Dynamical entanglement percolation with spatially correlated disorder}
	
	\author{Lorenzo Cirigliano}
	\email{lorenzo.cirigliano@uniroma1.it}
	\affiliation{The Abdus Salam International Centre for Theoretical Physics \href{https://ror.org/009gyvm78}{ICTP}, Trieste, Italy}
	\affiliation{Dipartimento di Fisica Universit\`a \href{https://ror.org/02be6w209}{La Sapienza}, I-00185 Rome, Italy}
	%
	\author{Valentina Brosco}
	\affiliation{Dipartimento di Fisica Universit\`a \href{https://ror.org/02be6w209}{La Sapienza}, I-00185 Rome, Italy}
	\altaffiliation{Istituto dei sistemi Complessi, Consiglio Nazionale delle Ricerche, Via dei Taurini, 19, Rome, Italy}
	\author{Claudio Castellano}
	\affiliation{\href{https://ror.org/05rcgef49}{Istituto dei Sistemi Complessi (ISC-CNR)}, Via dei Taurini 19, I-00185 Rome, Italy}
	%
	\author{Simone Felicetti}
	\affiliation{\href{https://ror.org/05rcgef49}{Istituto dei Sistemi Complessi (ISC-CNR)}, Via dei Taurini 19, I-00185 Rome, Italy}
	\author{Laura Pilozzi}
	\affiliation{\href{https://ror.org/05rcgef49}{Istituto dei Sistemi Complessi (ISC-CNR)}, Via dei Taurini 19, I-00185 Rome, Italy}
	%
	\author{Bernard van Heck}
	\affiliation{Dipartimento di Fisica Universit\`a \href{https://ror.org/02be6w209}{La Sapienza}, I-00185 Rome, Italy}
	\date{\today}
	
	\begin{abstract}
		The distribution of entanglement between the nodes of a quantum network plays a fundamental role in quantum information applications. %
		In this work, we investigate the dynamics of a network of qubits where each edge corresponds to an independent two-qubit interaction. %
		By applying tools from percolation theory, we study how long-range entanglement, obtained via singlet conversion and swapping from dynamically generated pairs, emerges across the network. We show that the interplay between unitary evolution and spatially correlated disorder leads to a non-standard percolation phenomenology, significantly richer than uniform bond percolation and featuring multi-valued hysteresis-like behaviour. %
		A two-colour correlated bond percolation model, whose phase diagram is determined via numerical simulations and a mean-field theory, fully elucidates the physics behind this phenomenon.%
	\end{abstract}
	
	\maketitle
	
\section{Introduction}

Quantum networks are at the basis of distributed quantum information protocols~\cite{ekert1991quantum,raussendorf2003measurement,meter2007communication}, enabling secure communication~\cite{wehner2018quantum,azuma2020tools,cirigliano2024optimal}, distributed quantum computing~\cite{spiller2006quantum,vanmeter2016the}, enhanced quantum sensing~\cite{komar2014a}, and clock synchronization~\cite{mckenzie2024clock}. Besides their relevance for quantum information applications, quantum networks are the ideal playground for modelling complex quantum devices and many-body systems~\cite{tura2014detecting,biamonte2019complex} and they can be utilized to represent the structure of complex quantum states~\cite{osterloh2002scaling,bianconi2013superconductor,walschaers2023emergent,mendessantos2024wave,bellomia2024quasilocal} and to benchmark the performance and efficiency of quantum processors and simulators~\cite{cavalcanti2011quantum,bhakuni2024diagnosing,andreoni2025network}.

Analogous to classical networks, an essential property of quantum networks is their long-range structure and connectivity. In classical complex networks, connectivity is typically determined by the underlying topology and link formation mechanisms~\cite{albert2002statistical}. In quantum networks, connectivity emerges from quantum correlations and can be engineered and controlled dynamically.
To enable quantum information processing on large scales, it is therefore crucial to develop reliable methods for establishing and maintaining long-range quantum connectivity. This entails the ability to distribute entanglement between distant nodes~\cite{brunner2014bell,das2018robust}, to manipulate quantum coherence and other quantum resources~\cite{adesso2016measures,streltsov2017colloquium}, and to generate multipartite entangled states such as Greenberger–Horne–Zeilinger (GHZ) and cluster states~\cite{shimizu2025simple,plodzien2025many}.

Entanglement percolation models~\cite{kieling2007percolation,acin2007entanglement,browne2008phase,cuquet2009entanglement,perseguers2010quantum,cuquet2011limited,perseguers2013distribution,romancino2026entanglement,romancino2026random} address the problem of creating entanglement between distant nodes when maximally entangled pairs are probabilistically established between nodes with a singlet-conversion probability $p$.
In one-dimensional networks, the probability of sharing a singlet between two distant nodes decays exponentially with their separation~\cite{acin2007entanglement}. In higher-dimensional structures, however, the number of possible connecting paths increases, giving rise to a percolation phenomenon: above a critical value of $p$, a giant connected cluster emerges that spans a finite fraction of the network. Once a path of singlets links two nodes, entanglement swapping allows the creation of a maximally entangled state between the end nodes, {\sl i.e.,} two arbitrary nodes can be connected by a singlet whenever they belong to the same cluster. As discussed in various works (see {\sl e.g.} Ref.~\cite{perseguers2013distribution}) the problem maps to a standard bond percolation~\cite{stauffer2018introduction}: when $p$ exceeds a critical threshold, entanglement percolates through the network, enabling quantum correlations to be distributed over arbitrarily large distances.

More recent advances have added layers of realism and sophistication to this picture.
They have for instance addressed path-based percolation where communication consumes resources~\cite{kim2024shortest,meng2025path,degirolamo2025percolation}, the importance of non-shortest paths in network resilience~\cite{hu2025unveiling}, entanglement distribution in complex network topologies~\cite{yang2022strong,brito2020statistical}, the effect of distributed quantum memories~\cite{meng2025quantum}, and alternative protocols that anticipate the percolation transition~\cite{shao2026entanglement}.

In all these approaches, entanglement spreads through the network by distributing entangled pairs among the nodes.
This perspective is natural within the context of quantum communication~\cite{wehner2018quantum,azuma2020tools}, for example in networks of optically linked stations. However, in many relevant cases such as Heisenberg spin chains~\cite{chiara2006entanglement,kim2013ballistic,eisert2015quantum,laflorencie2016quantum,hoshino2025entanglement}, entanglement is not simply distributed, but instead generated and modified by direct interactions between qubits. In such cases, entanglement becomes a dynamical resource.

In this work, we combine concepts from quantum network theory and percolation to investigate the dynamics of entanglement percolation in such interaction-driven quantum networks with spatially correlated disorder. The model we consider can be viewed as a time-dependent, coherent extension of the static entanglement-percolation scenario studied in previous works~\cite{acin2007entanglement,cuquet2009entanglement,cuquet2011limited,romancino2026entanglement,romancino2026random}. In these works, the singlet-conversion probabilities are assigned statically, either uniformly or as independent random variables.
In the latter case the problem has been shown to be equivalent, after averaging over the disorder, to uniform bond percolation~\cite{romancino2026entanglement}.
Our work goes beyond this setting in two respects: first, the singlet-conversion probabilities are generated dynamically and oscillate in time; second, they are spatially correlated, which, as we show, breaks the equivalence with uniform bond percolation.
The network is initialized in a product state at $t=0$ and then undergoes unitary evolution under qubit–qubit couplings for a finite time, after which the interactions are switched off. We then analyse the probability of generating long-range entanglement within a percolation framework.

We show that this dynamical setting can present a variety of interesting phenomena, in particular when unitary evolution occurs under a disordered network Hamiltonian with spatially correlated parameters.
In this case, the probability of establishing long-range entanglement in the network displays hysteresis as a function of time, so that the standard picture based on uniform bond percolation~\cite{acin2007entanglement} can no longer be applied.
Instead, we can understand this phenomenology using a two-colour inhomogeneous bond percolation model~\cite{chayes1987inhomogeneous,zhang1994a,kryven2019bond} with spatial constraints.
Given the simplicity of the underlying dynamics, these effects likely represent only a small part of the broader landscape that emerges when time-dependent processes are incorporated into the study of quantum networks.

The manuscript is structured as follows. In Sec.~\ref{sec:model} we describe the model in detail and recall the necessary notions of percolation theory. In Sec.~\ref{sec:uncorrelated}, we warm up by considering the case of a dynamical network with disorder but without correlations. In particular, we demonstrate some universal results on the percolating state of the network at long times, which indicate that the uniform bond percolation picture remains essentially unaltered. In Sec.~\ref{sec:spatial_correlations} we introduce spatial correlations into the picture, and reveal via controlled numerical experiments that they induce hysteresis in the time evolution of the network. In Sec.~\ref{sec:mapping} we describe the mapping to the two-colour bond percolation model, which provides a minimal setting to understand the occurrence of hysteresis.
\section{A dynamical model of entanglement percolation}
\label{sec:model}

\begin{figure}
	\centering
	\includegraphics[width=0.485\textwidth]{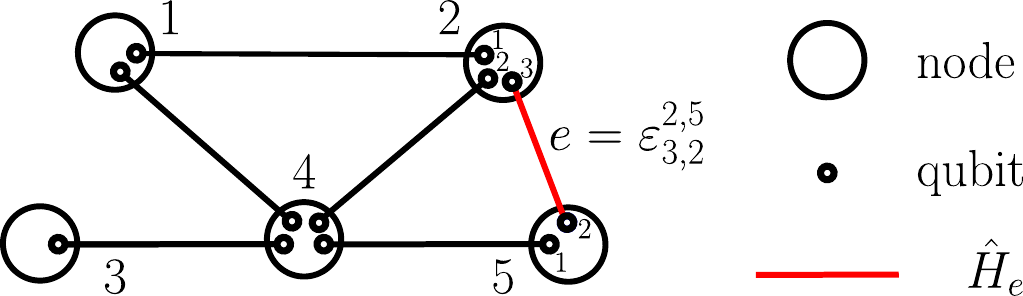}
	\caption{\textbf{Schematic representation of a quantum network.} We study a quantum network made of stations, forming the nodes of a graph, connected by edges that correspond to entangling interactions between pairs of qubits. Big circles denote stations, small circles represent qubits, lines indicate the presence of an interaction between two qubits. The interaction $\hat{H}_e$ between the qubits $j_A=3$ on station $A=2$ and $j_B=2$ on station $B=5$, represented by the edge $e=\varepsilon_{3,2}^{2,5}$, is highlighted in red. Note that each interaction involves two dedicated qubits, so that the number of qubits belonging to each station is equal to the degree of the corresponding node of the graph.}
	\label{fig:schematic_qnet}
\end{figure}

We consider a quantum network model with $N$ nodes representing stations, each with a given number of qubits. We label nodes with upper-case letters $A, B, \dots$ and denote as $j_A$ the $j$-th physical qubit at node $A$, with $j_A=1,2,\dots,n_A$. We assume that every qubit is associated with only one undirected edge, $\varepsilon^{A,B}_{j_{A}, j_{B}}$, of the graph, representing a dynamical connection between stations, {\sl i.e.,} the qubit $j_A$ on node $A$ interacting with the qubit $j_B$ placed on node $B$, see Figure~\ref{fig:schematic_qnet}.
This implies that, by construction, the number of qubits, $n_A$, in a station equals the degree, {\sl i.e.,} the number of connections, of the corresponding node $A$ in the graph.
The system's Hamiltonian has the form
\begin{equation}
	\label{eq:general_hamiltonian_L}
	\hat{H}=\frac{1}{2}\sum_{A < B } \sum_{j_A, j_B} J^{A,B}_{j_{A}, j_{B}}\hat{\sigma}_x^{A,j_A}\otimes \hat{\sigma}_x^{B,j_B},
\end{equation}
where the matrix $J$ defines the connections between the nodes and $\hat{\sigma}_x^{A,j_A}$ is the Pauli operator acting on the state of the qubit $j_A$ on station $A$. At time $t=0$ the system is initialized in a simple product state of the form 
\begin{equation}\label{eq:initial_product_state}
	\ket{\Psi_0}=\bigotimes_{A=1}^{N}\bigotimes_{ j_A=1}^{n_A}\ket{s}_{A,j_A},
\end{equation}
with $s \in \{0,1\}$, where $\ket{s}_{A,j_A}$ denotes the eigenstate of $\hat{\sigma}_z^{A,j_A}$ with eigenvalue $(-1)^s$. We choose the Hamiltonian~\eqref{eq:general_hamiltonian_L} as a simple model that can generate entanglement acting on any initial state of the form~\eqref{eq:initial_product_state}.

Given the structure of the system's Hamiltonian defined by Eq.~\eqref{eq:general_hamiltonian_L}, to describe the time evolution of the network we label as $e\equiv \varepsilon^{A,B}_{j_{A}, j_{B}}$ the generic element of the set $\mathcal{E}$ of all edges of the network, one for each pair of interacting qubits, and we introduce the entangling Hamiltonian $\hat{H}_e$ on the edge $e$ as
\begin{equation}
	\label{eq:edge_hamiltonian}
	\hat{H}_e = \frac{1}{2}\omega_e\,\hat{\sigma}_x^{A,j_A} \otimes \hat{\sigma}_x^{B,j_B} \,,
\end{equation}
where $\omega_e$ is the value of $J^{AB}_{j_{A} j_{B}}$ for the edge $e$. Note that in this notation, $\hat{H}=\sum_{e\in\mathcal{E}}\hat{H}_e$.

At time $t$, the state of the network can be expressed as a product of two-qubit states associated with each edge,
\begin{equation}
	\label{eq:time_evolution}
	\ket{\Psi(t)} =\bigotimes_{e \in \mathcal{E}} \ket{\psi_e(t)}\,,
\end{equation}
with $\ket{\psi_e(t)} = e^{-i\hat{H}_{e}t}\ket{\psi_e(0)}$, namely
\begin{align}
	\nonumber
	\ket{\psi_e(t)} &= \cos(\omega_et/2)\ket{s}_{A,j_A} \otimes\ket{s}_{B,j_B} \\
	&- i \sin(\omega_et/2) \,\hat{\sigma}_{x}^{A,j_A} \ket{s}_{A,j_A} \otimes \hat{\sigma}_{x}^{B,j_B}\ket{s}_{B,j_B}.
	\label{eq:time_evolution_single_edge}
\end{align}
The quantum state of the network is a product of two-qubit entangled states -- one {\sl per} edge $e$ -- each oscillating with frequency $\omega_e$. Indeed, as discussed in Appendix~\ref{appendix:schmidt}, applying the Schmidt decomposition to this two-qubit state yields
\begin{align}
	\nonumber
	\ket{\psi_e(t)} &= \sqrt{\lambda_e(t)}\ket{u}_{A,j_A}\otimes\ket{u}_{B,j_B} \\
	&+ \sqrt{1-\lambda_e(t)}\ket{v}_{A,j_A}\otimes\ket{v}_{B,j_B},
\end{align}
where $\ket{u}$ and $\ket{v}$ are some single-qubit states and the Schmidt coefficient is given by $\lambda_{e}(t) =\frac{1}{2}(1+ |\cos(\omega_e t)|)$.

\begin{figure}
	\centering
	\includegraphics[width=0.485\textwidth]{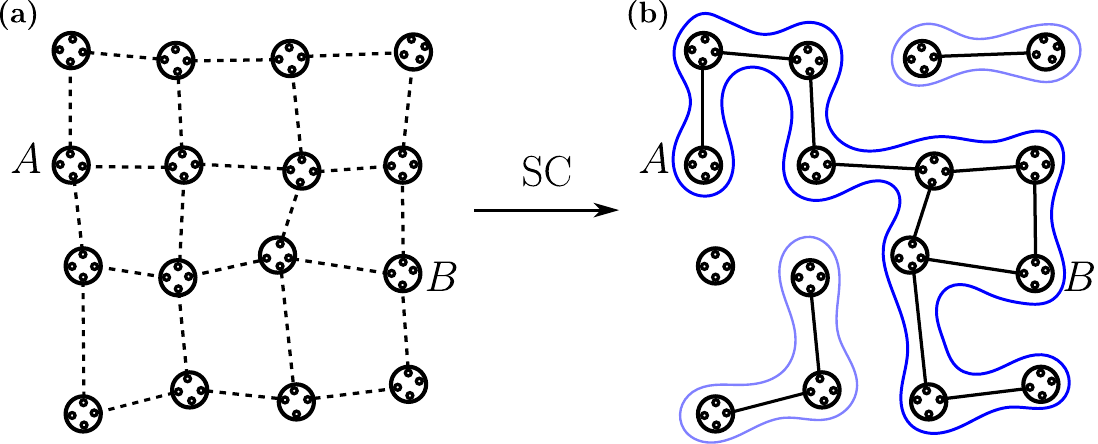}
	\caption{\textbf{Schematic representation of dynamical entanglement percolation.} (a) The dashed lines represent quantum states evolving as in Eq.~\eqref{eq:time_evolution_single_edge}. In this example, the network is a distortion of a square lattice, and spatial correlations are present between neighbouring edges. (b) Result of the singlet conversion (SC) process. At a certain time $t$, each edge state $\ket{\psi_e(t)}$ is converted to a maximally entangled state with the singlet-conversion probability $\phi_{e}(t)$ given in Eq.~\eqref{eq:activation_probability}. Successful conversions are represented with continuous lines. The maximally entangled connected components are highlighted in blue, with thicker lines for the largest component (a proxy for the giant component). The probability that $A$ and $B$ can share entanglement is related to the size of the giant component. In this case, they both belong to the giant component.}
	\label{fig:perturbed_lattice}
\end{figure}

Starting from this simple model of a dynamical quantum network, represented pictorially in Figure~\ref{fig:perturbed_lattice}(a), our purpose is to study the likelihood of sharing entanglement between arbitrarily distant qubits.
Following Ref.~\cite{acin2007entanglement}, we assume that at a given time $t$ it is possible to operate locally to attempt the conversion of $\ket{\psi_e(t)}$ into a maximally entangled state. The optimal singlet-conversion probability is~\cite{vidal1999entanglement}
\begin{equation}
	\label{eq:activation_probability}
	\phi_{e}(t) = \min\{1, 2(1-\lambda_{e}(t))\} = 1-|\cos(\omega_{e} t)|.
\end{equation}
After this operation has been carried out for each edge, one can use local entanglement swapping on maximally entangled adjacent edges in order to share entanglement among distant stations, provided that there is at least one path made of consecutive maximally entangled states connecting the corresponding nodes, see Figure~\ref{fig:perturbed_lattice}(b). 

Determining whether a maximally entangled path exists between two distant nodes is essentially equivalent to the study of connectivity properties of the network.
From a percolation perspective, we say that each edge is \textit{active} with probability $\phi_e(t)$ and we attempt to find a percolating path through the network.

Percolation theory offers guidance on the relevant quantities to study in order to characterize the entanglement of the network.
The first such quantity is the total fraction of maximally entangled states -- in the percolation jargon, the fraction of active edges
\begin{equation}
	\label{eq:average_fraction_active}
	p(t)=\frac{1}{E} \sum_{e \in \mathcal{E}} \phi_e(t)\,,
\end{equation}
where $E$ is the total number of edges in the network.

The second quantity, which we denote $P(t)$, is the fraction of nodes belonging to the largest component of nodes connected only via maximally-entangled states, see Figure~\ref{fig:perturbed_lattice}(b). Note that $P(t)$ is related to the probability that long-distance entanglement can be shared between two nodes picked at random: in the large $N$ limit, this probability is given by $P^2$. When the largest-component size scales with the size of the network, {\sl i.e.,} when $P(t)$ remains finite in the thermodynamic limit $N\to\infty$, we say that a giant component exists. In particular, if no giant component exists entanglement can be shared only locally. Even if the concept of giant component is formally well-defined only for $N \to \infty$, in practice the size of the largest connected component is a good proxy for it for finite $N$.

In percolation theory, the order parameter signalling that the network is in a percolating state is $P(p)$, that is the fraction of nodes belonging to the giant component as a function of the fraction of active edges~\cite{stauffer2018introduction}. In standard bond percolation with a fixed edge activation probability, $P$ and $p$ are static quantities and $P(p)$ is a well-defined single-valued function. We denote the order parameter of uniform bond percolation with $P_0(p)$. In contrast, we will see that in our dynamical model there may be multiple times along the time evolution characterized by the same $p(t)$ but by different $P(t)$. In this case, which occurs in the presence of correlations in the network parameters $\omega_e$, $P(p)$ may become a multi-valued function which deviates from the order parameter $P_0$.

\section{Effect of uncorrelated disorder}
\label{sec:uncorrelated}

\begin{figure*}
	\centering
	\includegraphics[width=0.95\textwidth]{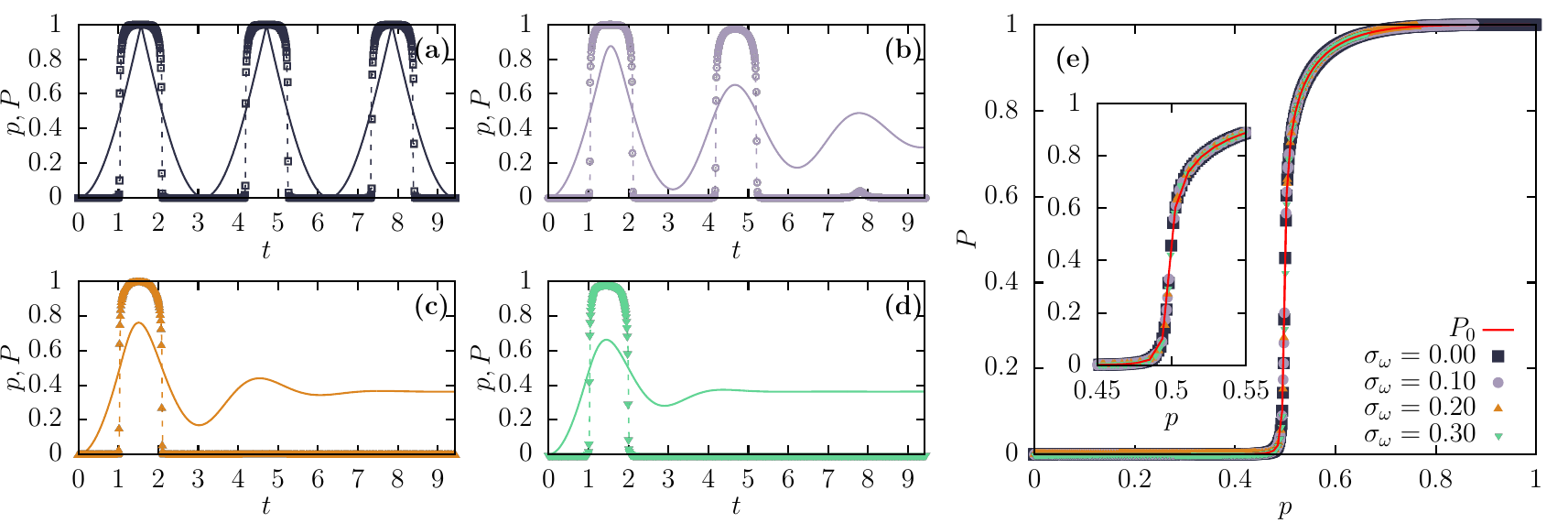}
	\caption{\textbf{Dynamical entanglement percolation observables on a two-dimensional square lattice with i.i.d. frequencies.} (a)-(d) The observables $p(t)$ (solid lines) and $P(t)$ (symbols and dashed lines) for independent Gaussian frequencies with mean $\overline{\omega}=1$ and (a) $\sigma_{\omega}=0$ (uniform frequencies), (b) $\sigma_{\omega}=0.1$, (c) $\sigma_{\omega}=0.2$, (d) $\sigma_{\omega}=0.3$. For uniform frequencies in panel (a), both $p$ and $P$ are periodic with period $T=\pi$. In all other cases, $p(t) \to p_\infty=1-2/\pi$ for large times. (e) Parametric plot of $P(p)$ using the observables $p(t)$ and $P(t)$ in panels (a)-(d). The effect of the dynamical evolution disappears when observing the order parameter $P(p)$, as all curves are equivalent and they all reproduce uniform bond percolation $P_0(p)$ (red line). The role of $t$ is just manifest in the fact that, depending on $\sigma_{\omega}$, not all values of $P$ can be reached. The inset in panel (e) shows the behaviour of $P(p)$ close to the bond percolation threshold $p_c=1/2$. All data perfectly overlap. For all plots, the system size is $L=10^3$, hence $N=10^6$. All results are averaged over $10$ realizations of the disorder, and over $100$ realizations of the edge activation process.}
	\label{fig:uncorrelated_noise}
\end{figure*}

In this section, we explore the consequences of the dynamical evolution of the network under the assumption that the frequencies $\omega_e$ are independent and identically distributed (i.i.d.).
We begin by noticing that if the frequencies are all equal, {\sl i.e.,} $\omega_e=\overline{\omega}$ for all edges, we fall back to the classical entanglement percolation of Ref.~\cite{acin2007entanglement}, except with a time-dependent singlet-conversion probability $\phi_e(t)=1-|\cos(\overline{\omega} t)|$, uniform across the entire network.
The fraction of active edges is therefore simply given by $p(t)=1-|\cos(\overline{\omega} t)|$, and the corresponding time evolution of $P(t)$ is also periodic in time, as shown in Figure~\ref{fig:uncorrelated_noise}(a) for the case of a square lattice of nodes.
The oscillations of $p(t)$ and $P(t)$ are always in phase, so the resulting curve $P(p)$ shown in Figure~\ref{fig:uncorrelated_noise}(e) is just the order parameter $P_0$ of bond percolation on a square lattice with uniform edge activation probability $p$~\cite{stauffer2018introduction}, a single-valued function.
The effect of the time evolution is only to move the state of the network back and forth along the curve $P_0(p)$ in a periodic fashion.

The situation becomes more interesting if we add some noise to the $\omega_e$. If the frequencies are random variables distributed according to some distribution $\mathcal{P}_{\omega}$, for large networks we may estimate the fraction of active edges as
\begin{equation}
	\label{eq:average_fraction_ME}
	p(t) = 1-\mathbb{E}\left[|\cos(\omega t) |\right],
\end{equation}
where $\mathbb{E}[\cdot]$ denotes the expectation value over $\mathcal{P}_{\omega}$. In Appendix~\ref{appendix:average_fraction_active}, we show that for any continuous distribution, at long times we reach a stationary state in which on average each edge is active with probability
\begin{equation}\label{eq:asymptotic_p}
	p_\infty=1-\frac{2}{\pi} \approx 0.363\,.
\end{equation}
This asymptotic result holds for any continuous probability density, as it is related to the large-argument behaviour of the Fourier transform of $\mathcal{P}_{\omega}$, see Appendix~\ref{appendix:average_fraction_active}. The detailed shape of the distribution only determines the pre-asymptotic behaviour of $p(t)$.

A stable giant component exists at long times if $p_\infty$ is larger than the standard bond percolation threshold $p_c$ of the network. In turn, $p_c$ depends on the topology of the interactions in the network.
For example, for the triangular lattice $p_c = 2 \sin(\pi/18)\approx 0.347$~\cite{stauffer2018introduction} and a giant component exists asymptotically.
On the other hand, for the square lattice $p_c=1/2$~\cite{stauffer2018introduction}, so long-distance entanglement cannot be established at long times.
However, depending on the shape of the distribution, there may be intermediate times $t$ such that $p(t)>p_c$, in which case a transient giant component occurs. 

These results are corroborated numerically in Figure~\ref{fig:uncorrelated_noise} for a square lattice with i.i.d. frequencies with a Gaussian distribution with average value $\overline{\omega}$ and standard deviation $\sigma_{\omega}$.
In this case, the asymptotic behaviour is reached when $\sigma_{\omega} t\gg 1$.
As seen in panels (a)-(d) of Figure~\ref{fig:uncorrelated_noise}, for increasing values of $\sigma_{\omega}$ the oscillations in $p(t)$ saturate more quickly to the asymptotic value.

While in the disordered case the time evolution of $p$ and $P$ is no longer periodic, their damped oscillations remain in phase as they did in the disorder-free case.
The plot of $P$ versus $p$ in Figure~\ref{fig:uncorrelated_noise}(e) shows a collapse over a single curve, even when including data points from multiple oscillations and from different values of $\sigma_{\omega}$.

In other words, although the system evolves dynamically in a way that depends on the disorder realization, at each time the values of $p(t)$ and $P(t)$ always fall on the curve $P_0(p)$ provided by the order parameter of uniform bond percolation.
The network now performs damped oscillations back and forth along this curve and always ends up at the asymptotic point $P(p_\infty)=P_0(p_{\infty})$. These empirical observations are supported by the theory developed in Ref.~\cite{romancino2026entanglement}, where it has been shown analytically that for random independent singlet-conversion probabilities, the process is equivalent, after averaging over the disorder, to uniform bond percolation. In the next Section, we show that the inclusion of spatial correlations in the network changes this picture qualitatively, exhibiting hysteresis.
\section{Adding spatial correlations}
\label{sec:spatial_correlations}

Instead of considering independent random frequencies for different edges of the network, let us introduce correlations between frequencies of edges connected to the same node. A general way to do so is to first assign a weight $g_A$ to each node $A$ of the quantum network and assume that the frequency associated with the edge $\varepsilon^{A,B}_{j_A, j_B}$ is a function of $g_A$ and $g_B$, {\sl i.e.,}, $\omega_{e}=f(g_A,g_B)$. Here $f$ is a symmetric function of its two arguments, which we assume to be the same for every edge of the network. The dependence of the frequency $\omega_e$ on $g_A$ and $g_B$ introduces correlations between frequencies of neighbouring edges.
The weight $g_A$ may be related to physical properties of the station $A$, for example its position and the nature of the connection with the other nodes.

\begin{figure*}
	\centering
	\includegraphics[width=0.95\textwidth]{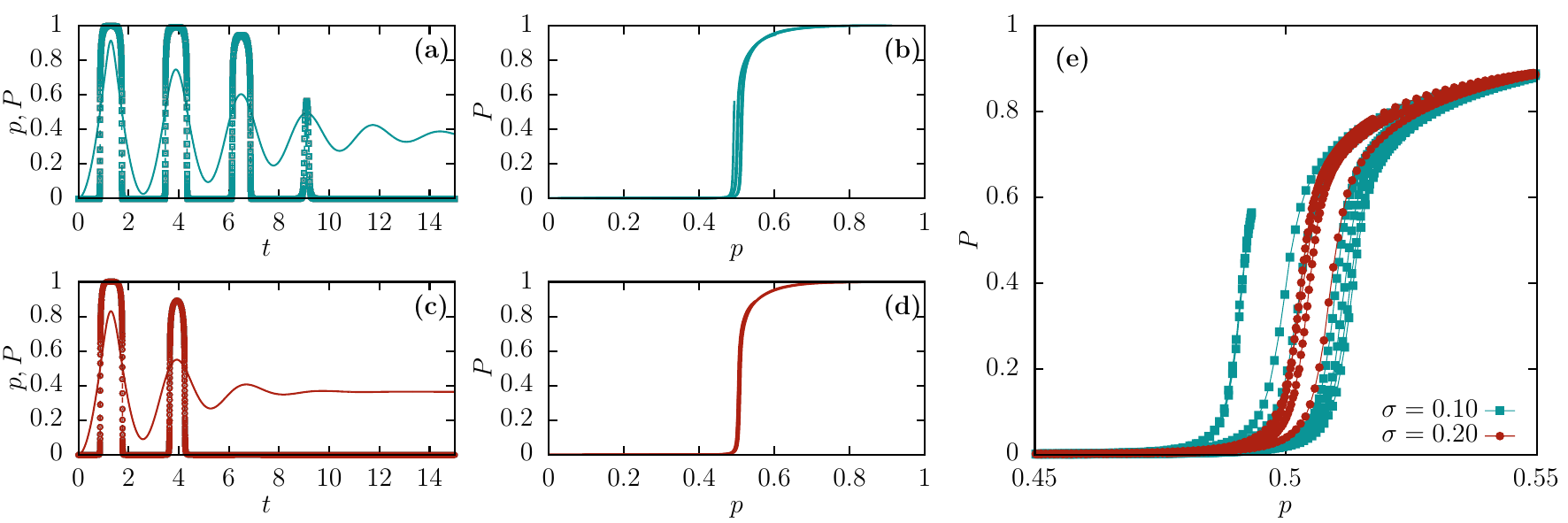}
	\caption{\textbf{Dynamical entanglement percolation observables on a perturbed square lattice with exponentially decaying frequencies.} Results of numerical simulations on a perturbed square lattice with noise strength $\sigma$ and frequencies depending on the distances as $\omega(d)=\omega_0 e^{-d/\ell}$. We set $\omega_0=2$, $\ell=2$. (a) The observables $p(t)$ (solid lines) and $P(t)$ (symbols) for $\sigma=0.1$, and (b) the corresponding parametric plot of $P(p)$. (c)-(d) As in (a)-(b) but for $\sigma=0.2$. In both cases, the dynamics converges to the asymptotic value $p_\infty=1-2/\pi$, after a pre-asymptotic oscillatory behaviour which depends on the noise amplitude. (e) The behaviour of the observables $P(p)$ as in panels (b) and (d) close to the uniform percolation threshold on the square lattice $p_c=1/2$~\cite{stauffer2018introduction}. In contrast to the case of independent random frequencies, $P$ clearly depends on $\sigma$, and the order parameter $P(p)$ is no longer the same as that of uniform bond percolation.}
	\label{fig:correlated_noise_exponential}
\end{figure*}

Let us provide a practical instance of this procedure which is physically transparent and suitable for numerical study.
We start with a two-dimensional square lattice of nodes with unit lattice spacing.
We then randomly perturb the position of each node with a Gaussian noise $(\delta x,\delta y)$, where $\delta x$ and $\delta y$ are i.i.d. Gaussian variables with zero mean and variance $\sigma^2$, see Figure~\ref{fig:perturbed_lattice}(a).
As a result, the Euclidean distances $d$ between two perturbed neighbouring nodes of the lattice are correlated, in a way detailed in Appendices~\ref{appendix:distances} and~\ref{appendix:theta}.
The frequency $\omega_e$ of an edge connecting two stations is then chosen to be a deterministic function of the distance between the two nodes: $\omega_e\equiv \omega(d)$.

In Figure~\ref{fig:correlated_noise_exponential} we report results of numerical simulations of the dynamical evolution of a network on the perturbed square lattice with frequencies that decay exponentially over a length scale $\ell$, $\omega(d)=\omega_0\,e^{-d/\ell}$.
As in the case without correlations, at long times $p(t)$ tends to $p_\infty=1-2/\pi$, and the asymptotic limit is reached faster when disorder is stronger. Furthermore, as before, $P(t)$ becomes finite when $p(t)$ is large enough.
However, in contrast to the case of i.i.d. frequencies, the $P(p)$ trajectory in panel (e) shows many branches: $P(t)$ has become a multi-valued function of $p(t)$.
This hysteretic behaviour is the manifest effect of local correlations in the frequencies induced by the spatial constraint.

The finite-time behaviour in Figure~\ref{fig:correlated_noise_exponential} depends on the specific choice of the function $\omega(d)$, which determines the frequency distribution of the network.
In order to isolate the effect of correlations, we strip down our frequency assignment procedure to a threshold model in which frequencies can take one out of only two different values
\begin{equation}
	\label{eq:frequencies_threshold}
	\omega(d) =
	\begin{cases}
		\Omega_1 &  \textrm{if}\quad  d>\ell\,,\\
		\Omega_2 &  \textrm{if}\quad d<\ell\,.
	\end{cases}
\end{equation}
Note that with no loss of generality we can set $\Omega_1=1$, or equivalently rescale $t$ by $\Omega_1$, and use only one parameter $\widetilde{\Omega}=\Omega_2/\Omega_1$. Because of the presence of the geometric constraint, the frequencies of edges having a common node are correlated.
As shown in Appendix~\ref{appendix:theta}, for the two-frequency model considered here, it is possible to fully characterize such correlations using only three probabilities which are determined by the model parameters $\sigma$ and $\ell$: (i) the probability $\eta$ that $d>\ell$; (ii) the probability that two adjacent edges along the same direction are both shorter than $\ell$; (iii) the probability that two adjacent edges along orthogonal directions are both shorter than $\ell$. As extensively discussed in Appendix~\ref{appendix:theta}, for $\sigma < 0.3$ and $\ell \approx 1$, the frequency distribution is balanced ($\eta \approx 1/2$): the number of edges with $\omega_e=\Omega_1$ and $\omega_e=\Omega_2$ is approximately the same. Furthermore, in this regime the frequencies are strongly anti-correlated along the same direction, either vertical or horizontal, and uncorrelated along orthogonal directions.

\begin{figure*}
	\centering
	\includegraphics[width=0.95\textwidth]{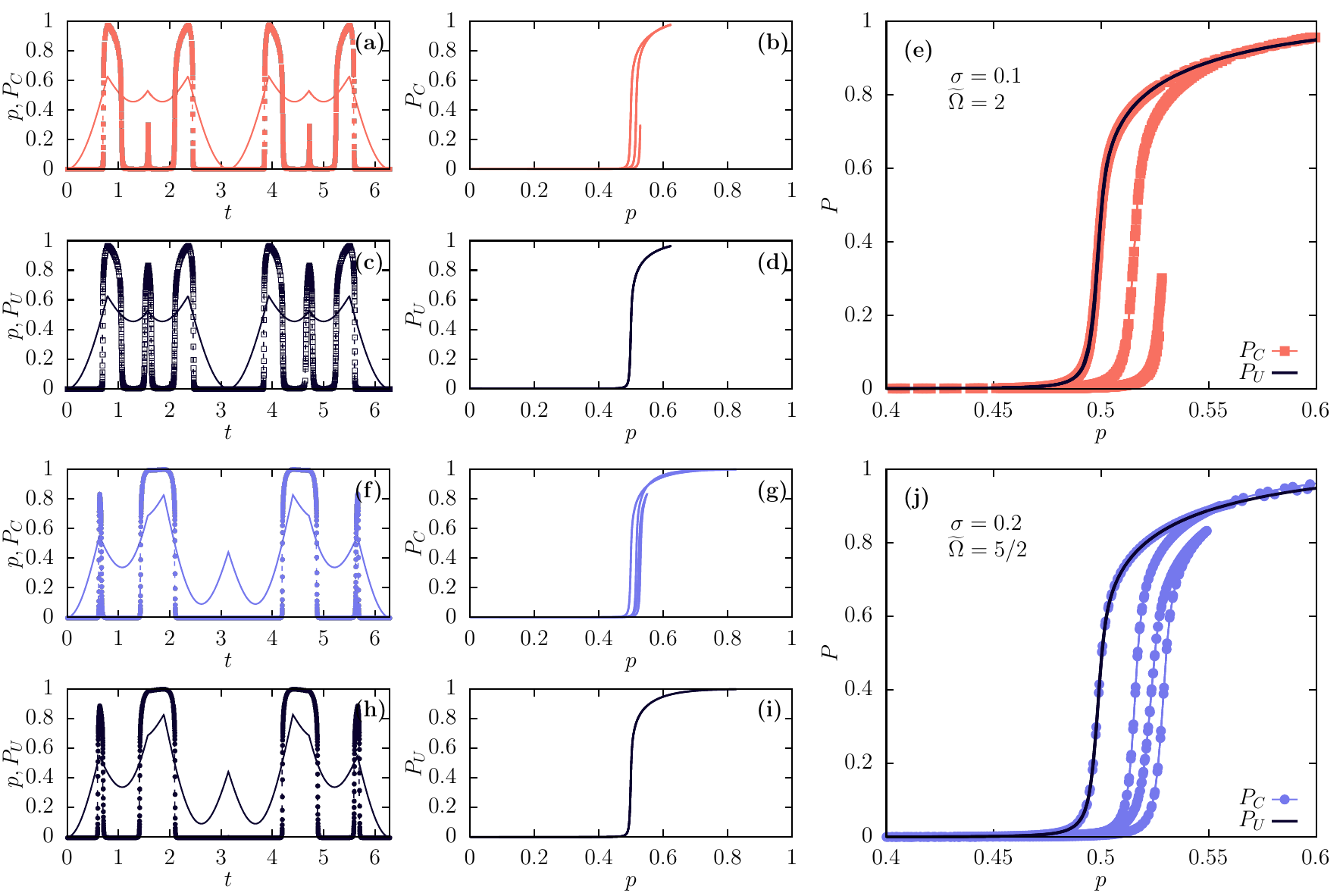}
	\caption{\textbf{Dynamical entanglement percolation observables on a perturbed square lattice with correlated Bernoulli frequencies.} Results of numerical simulations on a perturbed square lattice with noise strength $\sigma$ and frequencies depending on the distances as in Eq.~\eqref{eq:frequencies_threshold}. (a) The observables $p(t)$ (solid lines) and $P_C(t)$ (symbols and dashed lines) in the correlated case for $\sigma=0.1$, $\widetilde{\Omega}=2$, $\ell=1$, and (b) the corresponding parametric plot of $P_C(p)$. (c)-(d) As in (a)-(b) but for the uncorrelated case. (e) The behaviour of the observables $P_C(p)$ and $P_U(p)$ as in panels (b) and (d), respectively, close to the uniform percolation threshold on the square lattice $p_c=1/2$~\cite{stauffer2018introduction}.
		(f)-(j) As in (a)-(e) but for $\sigma=0.2$, $\widetilde{\Omega}=5/2$, $\ell=1$.
		The comparison between the correlated and the uncorrelated scenarios shows that the order parameters $P_C$ clearly depend on the parameters $\sigma$ and $\widetilde{\Omega}$, exhibiting a nontrivial behaviour with coexisting branches.}
	\label{fig:correlated_noise_threshold}
\end{figure*}

In Figure~\ref{fig:correlated_noise_threshold} we report the results of numerical simulations of this threshold model for some values of $\sigma$, $\widetilde{\Omega}$, and $\ell=1$.
To isolate the role played by correlations, for each simulation we run a parallel simulation in which, after fixing the realization of the disorder, we reshuffle the frequencies and reassign them to random edges.
In this way we can compare two networks with exactly the same frequency distribution, but one with correlations and one without. 
In Figure~\ref{fig:correlated_noise_threshold}, we use the subscripts $C$ and $U$ on the observables to denote the correlated case and the uncorrelated case, respectively.

In either case, the frequencies follow a discrete Bernoulli distribution. As shown in Appendix~\ref{appendix:average_fraction_active}, the expected fraction of active edges is in this case
\begin{equation}
	\label{eq:average_fraction_bernoulli}
	p(t) = 1 - \eta |\cos(\Omega_1 t)| - (1-\eta)|\cos(\Omega_2 t)|.
\end{equation}
It does not have an asymptotic value, but it undergoes periodic oscillations with a period which depends on the frequency ratio $\widetilde{\Omega}$, and which becomes infinite (the evolution is quasi-periodic) if $\widetilde{\Omega}$ is irrational. Importantly, the curve $p(t)$ does not depend on the presence of correlations between neighbouring edges.

The effect of correlations arises when inspecting $P(t)$.
In the uncorrelated case, $P_{U}(t)$ follows the evolution of $p(t)$, and $P_U(p)$ is a well-defined function, as discussed in Sec.~\ref{sec:uncorrelated} and in agreement with Ref.~\cite{romancino2026entanglement}. In the correlated case, instead, $P_C(t)$ shows the same type of multi-valued behaviour already observed in Figure~\ref{fig:correlated_noise_exponential} for a continuous distribution with correlations, see Figure~\ref{fig:correlated_noise_threshold}(e) and (j).
The value of the order parameter at time $t$ thus depends not only on the instantaneous value of $p$, but also on the previous history of the process $p(t)$, leading to hysteresis.
Note also that $P_C(t)$ rises from zero at different values of $p(t)$ for different branches of the hysteretic time evolution: in other words, the concept of a percolation threshold becomes blurred as different thresholds exist.
Note that if $\widetilde{\Omega}$ is a rational number $l/m$ with $l,m$ coprime, the time evolution is periodic with period $T=m \pi/\Omega_1$. In this case there is a finite set of percolation thresholds.
\section{Mapping to two-colour bond percolation}
\label{sec:mapping}

\begin{figure*}
	\centering
	\includegraphics[width=0.95\textwidth]{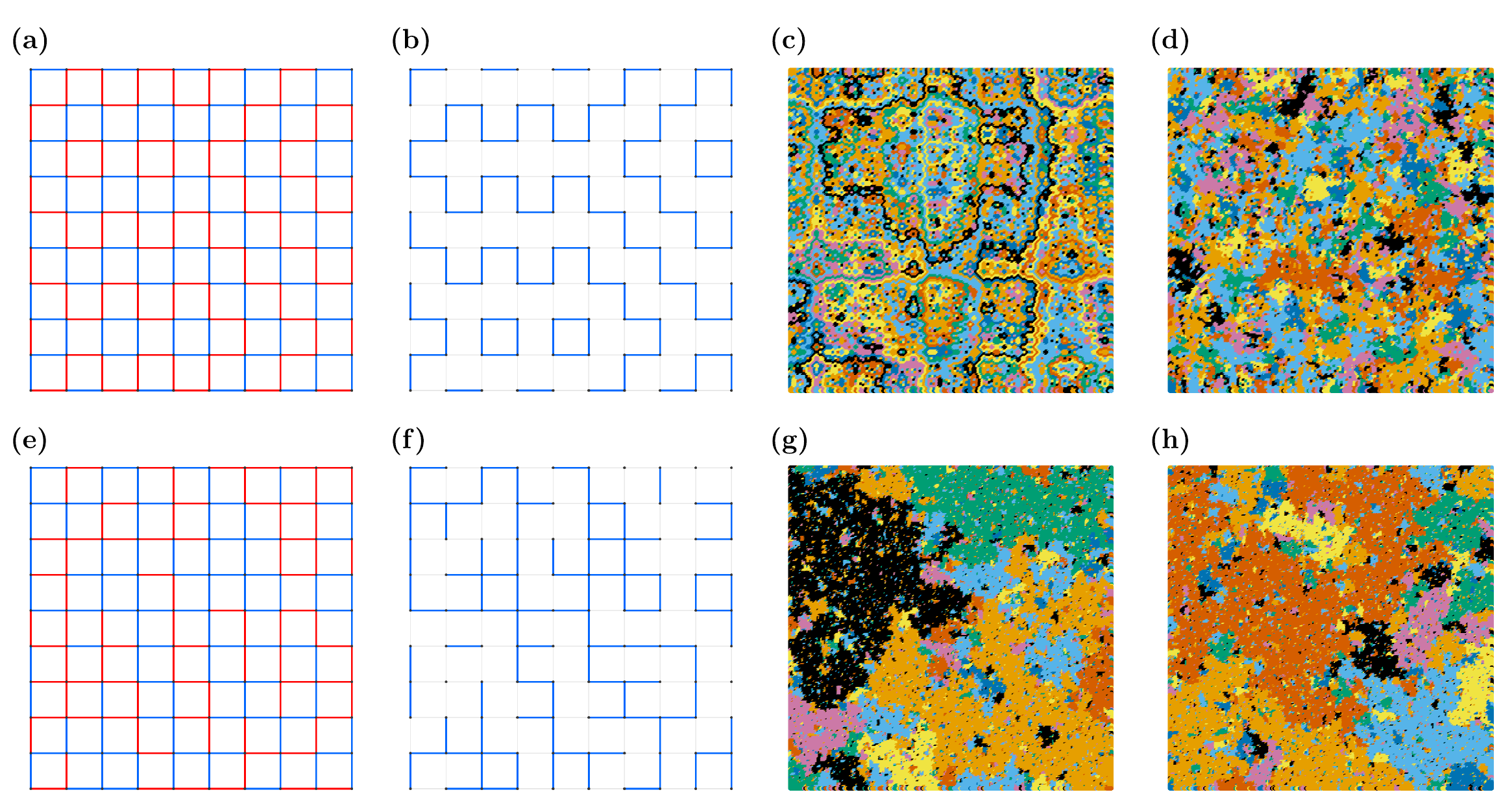}
	\caption{\textbf{Two-colour bond percolation on a square lattice.} (a) A configuration of the two-colour bond percolation model on a $10 \times 10$ square lattice. In (b) only blue edges are active, showing the presence of structural patterns. Snapshots of the connected components (coloured) on a $256 \times 256$ lattice for $p=0.5$, using (c) $\phi_1=0$, $\phi_2=1$, and (d) $\phi_1=0.1, \phi_2=0.9$. Even though the fraction of active edges is the same, the two configurations are markedly distinct because of the strong spatial constraints. Panels (e)-(h) are as in (a)-(d) but for a random placement of the two colours, without the local constraints. The configurations in (g) and (h) are statistically equivalent.}
	\label{fig:two_colour_model}
\end{figure*}

The results in the previous section demonstrate that dynamical entanglement percolation does not map to uniform bond percolation, because of the interplay between non-monotonic time evolution and correlated disorder.
We now show that the more complex behaviour is qualitatively captured by a two-colour inhomogeneous bond percolation model~\cite{chayes1987inhomogeneous,zhang1994a,scullard2010critical} with a specific type of spatial constraints.
We assume that the edges of a regular square lattice are of two colours, say red and blue. We also assume that half of the edges are of type 1 (red) and active with probability $\phi_1$, and half are of type 2 (blue) and active with probability $\phi_2$. The average fraction of active edges is $p=(\phi_1+\phi_2)/2$, and for this model we denote by $S(\phi_1,\phi_2)$ the fraction of nodes in the giant component.
To mimic the spatial correlation patterns observed in perturbed lattices, we assume that along each direction edge colours are alternating, repeating a red/blue pattern, and that the colours of two adjacent orthogonal edges are independent. As shown in Figure~\ref{fig:two_colour_model}(a)-(b), this constraint leads to large-scale structural patterns. These are absent in samples where the red and blue edges occur with the same proportions but in randomly reshuffled positions, as shown in Figure~\ref{fig:two_colour_model}(e)-(f).

Whether or not $S$ can be expressed in terms of $p$ corresponds to asking whether or not the curves $\phi_1+\phi_2=2p$ are level curves of $S$. If that is the case, we can write $P(p)=S(\phi_1,2p-\phi_1)$ which is independent of $\phi_1$. We expect that the spatial constraints producing the alternating pattern in the bond colours affect the behaviour of the order parameter. For instance, Figure~\ref{fig:two_colour_model}(c) and (d) show two configurations on the same two-colour lattice and for the same average fraction of active edges $p=1/2$ but for (c) $\phi_1=0$, $\phi_2=1$, and (d) $\phi_1=0.1$, $\phi_2=0.9$. The presence of visually different spatial patterns for the same value of $p$ indicates that the cluster configurations also depend on which edges are active, not only on how many edges are active. This effect is instead absent if one considers a randomized placement of the two colours, as shown in Figure~\ref{fig:two_colour_model}(e)-(h).

\begin{figure*}
	\centering
	\includegraphics[width=0.95\textwidth]{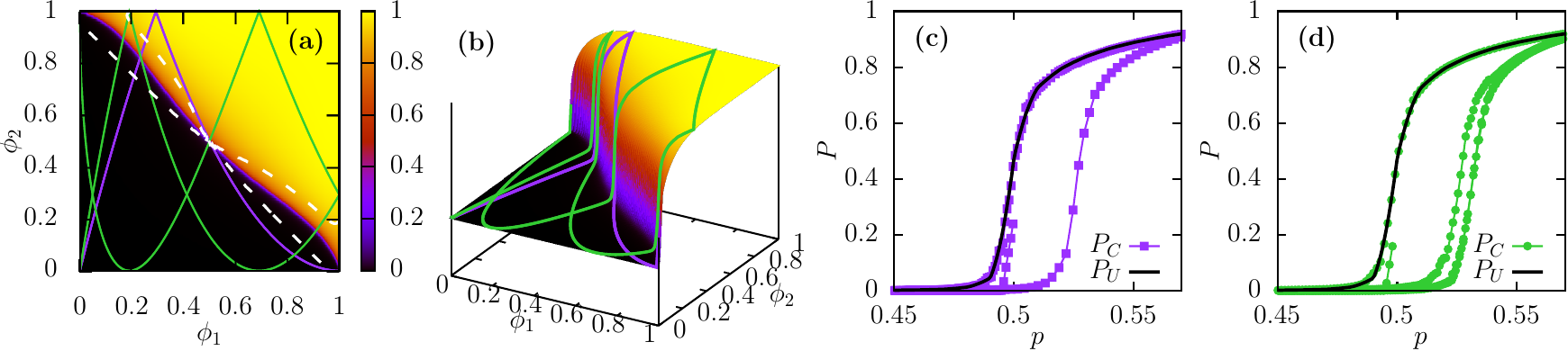}
	\caption{\textbf{Phase diagram of two-colour bond percolation on a square lattice.} (a) The phase diagram of $S_C$ in the $\phi_1-\phi_2$ plane. The continuous lines correspond to a dynamical evolution of $\phi_1$ and $\phi_2$ as in Eq.~\eqref{eq:dynamical_activation}, for $\widetilde{\Omega}=2$ (purple) and $\widetilde{\Omega}=5/2$ (green). The white dashed lines bound the region where $|S_U-S_C|>10^{-2}$. (b) 3D plot showing the surface $S(\phi_1,\phi_2)$. Moving on this surface following the purple and green curves we recover $P(t)$. (c) Parametric plot for the order parameter $P_C(p)$ (symbols) and $P_U(p)$ (black line) for $\widetilde{\Omega}=2$ (purple). (d) Same as in (c) but for $\widetilde{\Omega}=5/2$. All numerical simulations are performed on square lattices with $N=10^6$ nodes, averaged over $100$ independent realizations.}
	\label{fig:two_colour_phase}
\end{figure*}

We determine the phase diagram $S(\phi_1,\phi_2)$ by means of extensive numerical simulations on large lattices, both for the alternating pattern ($S_C$) and for the randomly reshuffled lattice ($S_U$). Results are presented in Figure~\ref{fig:two_colour_phase}(a). The closer to the percolating line, the more visible is the effect of the spatial constraints: the white dashed line bounds the region where $|S_U-S_C|>10^{-2}$.

At this point, the role of the dynamics is straightforward to understand. Edge activation probabilities changing in time can be represented by a curve $\gamma(t)=(\phi_1(t), \phi_2(t))$ in the parameter space, see for instance the two lines in Figure~\ref{fig:two_colour_phase}(a) corresponding to time evolutions of the form
\begin{equation}
	\label{eq:dynamical_activation}
	\phi_1(t)=1-|\cos(t)|,\quad\phi_2(t)=1-|\cos(\widetilde{\Omega} t)|,
\end{equation}
with $\widetilde{\Omega}=2$ (purple) and $\widetilde{\Omega}=5/2$ (green). Note that this choice of $\gamma(t)$ is meant to mimic the singlet-conversion probabilities given in Eq.~\eqref{eq:activation_probability}. However, any function $\gamma: \mathbb{R} \to [0,1] \times [0,1]$ can be used to model the time evolution of $\phi_1$ and $\phi_2$. The order parameter $P(t)$ is obtained by moving on the surface $S$ along the curve $\gamma$, that is $P_{C,U}(t)=S_{C,U}(\gamma(t))$, see Figure~\ref{fig:two_colour_phase}(b). Finally, the order parameter $P(p)$ is plotted in Figure~\ref{fig:two_colour_phase}(c) and (d), and it clearly shows the same hysteretic behaviour observed for the case of dynamical entanglement percolation in perturbed lattices. Such a multi-branch behaviour is the combined effect of (i) non-monotonic dynamical evolution of the edge activation probabilities, and (ii) strong spatial constraints. 

Remarkably, for both $\widetilde{\Omega}=2$ and $\widetilde{\Omega}=5/2$ there exists one branch of $P_C$ that overlaps with the order parameter $P_U$ of uniform bond percolation. This happens because both curves pass through the point $\phi_1=\phi_2=1/2$, where the distinction between the two colours disappears and the alternating pattern does not play any role.\footnote{%
	One may wonder if this happens for any $\widetilde{\Omega}$. Solving $\phi_1(t^*)=1/2$ gives $t^*=j\pi \pm \pi/3$ with $j \in \mathbb{Z}$, and imposing $\phi_2(t^*)=1/2$ gives $\widetilde{\Omega} t^*=k\pi \pm \pi/3$. Hence $\widetilde{\Omega} = \frac{3 k \pm 1}{3 j \pm 1}$. Writing $\widetilde{\Omega}=l/m$, with $l, m$ coprime, this is possible if and only if neither $l$ nor $m$ is a multiple of $3$. Then for rational $\widetilde{\Omega}$ satisfying this condition, $P_C$ has one branch, among the possibly many, that is equivalent to the function $P_U$ close to $p=1/2$.} 
Close to this point, as one can see from the white dashed lines in Figure~\ref{fig:two_colour_phase}(a), the two order parameters are almost equal.

Finally, we argue that this hysteretic behaviour is a truly thermodynamic effect that does not disappear in the infinite-size limit. To support this point, in Appendix~\ref{appendix:mean_field} we derive the mean-field solution for the two-colour bond percolation model, recovering a scenario qualitatively equivalent to what we observe here.
\section{Discussion}
\label{sec:discussion}

Entanglement percolation~\cite{kieling2007percolation,acin2007entanglement,browne2008phase,cuquet2009entanglement,perseguers2010quantum,cuquet2011limited,perseguers2013distribution,romancino2026entanglement,romancino2026random}
determines the minimum amount of entanglement required between nearest-neighbour nodes sharing an entangled qubit pair in order to establish a perfect long-distance quantum channel across a quantum network.
Starting from the picture in which network nodes correspond to physical devices or stations hosting qubits, we extend the standard model of entanglement percolation to account for correlated singlet-conversion probabilities across edges connected to the same node. In particular, motivated by the observation that spatial constraints and other physical limitations shape the topology of quantum networks -- similarly to the considerations in Ref.~\cite{cirigliano2024optimal} for the optimization of quantum key-distribution networks -- we introduce correlations that depend on the distance between nodes, going beyond the conventional independent-edge assumption. Furthermore, we consider scenarios in which entanglement between nodes is dynamically generated via a two-qubit interaction and oscillates in time due to the underlying Hamiltonian evolution. We analyse how this temporal behaviour affects the emergence of long-range entanglement when oscillation frequencies vary locally across the network.

Our numerical and analytical study leads to the following general findings. %
First, when the frequency distribution is continuous, the network evolves to a steady state in which each edge can be converted into a maximally entangled pair with average probability $1-2/\pi$. Whether this corresponds to a percolating phase depends on the network topology. %
Second, when frequencies are independent, the percolation properties are accurately captured by the usual bond-percolation order parameter, in close analogy with the static case considered in previous works, and in agreement with the analytical results of Ref.~\cite{romancino2026entanglement}. %
Third, when frequencies are correlated, the order parameter becomes a multi-valued function of the fraction of active edges, showing hysteretic behaviour as the network oscillates in time between percolating and non-percolating phases.
Finally, we show that a two-colour inhomogeneous bond-percolation model with spatial constraints provides a minimal theoretical framework capturing the key features of the correlated regime.

Our model is simple because the entanglement does not spread during the unitary evolution: the quantum state of the network is a simple tensor product of two-qubit states associated with each edge of the network.
Nevertheless, the resulting behaviour is rather complex from a percolation theory perspective, going qualitatively beyond that of standard entanglement percolation. 
A challenging and interesting task would be to study the effect of entanglement spreading during unitary evolution, {\sl e.g.} through the action of local interaction terms at each node of the network.

\section*{Acknowledgments}
This work was co-funded by Project PNRR MUR
PE 0000023-NQSTI  and  Project PNRR MUR project CN
00000013-ICSC.
C.C. acknowledges the PRIN project No. 20223W2JKJ “WECARE”, CUP
B53D23003880006, financed by the Italian Ministry of University and
Research (MUR), Piano Nazionale Di Ripresa e Resilienza (PNRR),
Missione 4 “Istruzione e Ricerca” - Componente C2 Investimento 1.1,
funded by the European Union - NextGenerationEU.

\section*{Data availability} 
All the data generated for the numerical studies reported in this work can be reproduced using the code available at~\cite{cirigliano2026github}.
	
	%
	\bibliography{references}
	
	\onecolumngrid
	\appendix
	\section{Schmidt decomposition}
	\label{appendix:schmidt}
	
	Given an arbitrary two-qubit state $\ket{\psi}=\sum_{i,j}\Psi_{ij}\ket{i}\otimes\ket{j}\in\mathcal{H}_A\otimes\mathcal{H}_B$, the Schmidt decomposition~\cite{nielsen2010quantum} tells us that there exist orthonormal bases $\{\ket{A_n}\}$ of $\mathcal{H}_A$ and $\{\ket{B_n}\}$ of $\mathcal{H}_B$ such that
	\begin{equation}
		\ket{\psi}=\sum_{n} \sqrt{\alpha_n} \ket{A_n}\otimes \ket{B_n}.
	\end{equation}
	In particular, this is precisely the singular value decomposition of the matrix $\Psi$~\cite{horn2012matrix} formed by the coefficients $\Psi_{ij}$. Thus the coefficients $\alpha_n$ are the (real, non-negative) eigenvalues of the matrix $\Psi^{\dagger}\Psi$. For a system of two qubits this matrix is a $2 \times 2$ Hermitian, positive semi-definite matrix, and we can express its eigenvalues in terms of trace and determinant. A straightforward computation gives us
	\begin{equation}
		\Psi^{\dagger}\Psi = \begin{bmatrix}
			|\Psi_{00}|^2 + |\Psi_{10}|^2 & \Psi_{00}^{*}\Psi_{01}+\Psi_{10}^{*}\Psi_{11}\\
			\Psi_{00}\Psi_{01}^{*}+\Psi_{10}\Psi_{11}^{*}  & |\Psi_{01}|^2 + |\Psi_{11}|^2
		\end{bmatrix},
	\end{equation}
	from which we get $\tr{\Psi^{\dagger}\Psi}=1$ and
	\begin{align}
		\det{\Psi^{\dagger}\Psi} &= |\det{\Psi}|^2 =  |\Psi_{00}|^2|\Psi_{11}|^2 + |\Psi_{01}|^2|\Psi_{10}|^2 - 2\Re\left\{\Psi_{00}\Psi_{01}^{*}\Psi_{10}^{*}\Psi_{11}\right\}.
	\end{align}
	Hence we have
	\begin{align}
		\alpha_{0} &= \frac{1 + \sqrt{1-4|\det{\Psi}|^2}}{2}, \\
		\alpha_{1} &= \frac{1 - \sqrt{1-4|\det{\Psi}|^2}}{2}.
	\end{align}
	Note that $\alpha_1=1-\alpha_0$, since the trace equals $1$, and we need only one Schmidt coefficient which we denote by $\lambda= \alpha_0$. States with $|\det{\Psi}|=1/2$ are maximally entangled states, while states with $|\det{\Psi}|=0$ are product states.
	
	\section{Average fraction of maximally entangled states}
	\label{appendix:average_fraction_active}
	
	In this appendix we compute the quantity $p(t)$ given in Eq.~\eqref{eq:average_fraction_active}. Let us assume that the frequencies are random variables following the probability density $\mathcal{P}_{\omega}$. We do not need the frequencies to be independent; we only require that the marginal density is the same for all of them. We can write
	\begin{equation}
		p(t) = 1 - \mathbb{E}\left[|\cos(\omega t)| \right] = 1-\int_{-\infty}^{+\infty} dx \mathcal{P}_{\omega}(x)|\cos(x t)| .
	\end{equation}
	
	To compute the integral, we set $y = x t$ and expand $|\cos y|$ in a Fourier series
	\begin{equation}
		|\cos(y)|=\frac{2}{\pi} - \frac{4}{\pi} \sum_{k=1}^{\infty}\frac{(-1)^k}{4k^2-1}\cos(2ky),
	\end{equation}
	and we get
	\begin{equation}
		\mathbb{E}\left[|\cos(\omega t)| \right] = \frac{2}{\pi} - \frac{4}{\pi} \sum_{k=1}^{\infty}\frac{(-1)^k}{4k^2-1} \mathbb{E}\left[\cos(2k\omega t) \right] .
	\end{equation}
	Writing $\cos(2k \omega t)=\Re\left[e^{i2k\omega t} \right]$ we finally get
	\begin{equation}
		p(t) = 1-\frac{2}{\pi} + \frac{4}{\pi} \sum_{k=1}^{\infty}\frac{(-1)^k}{4k^2-1} \Re \left[F(2 k t)\right],
	\end{equation}
	where $F(q)$ is the Fourier transform of the probability density $\mathcal{P}_{\omega}$
	\begin{equation}
		F(q)=\mathbb{E}\left[e^{i q \omega} \right] = \int dx \mathcal{P}_{\omega}(x)e^{i q x}.
	\end{equation}
	The problem of computing $p(t)$ is then reduced to computing the Fourier transform of $\mathcal{P}_{\omega}$. In particular, the asymptotic behaviour for large $t$ is captured by the asymptotic behaviour of $F(q)$ for large $q$. For any absolutely continuous $\mathcal{P}_{\omega}$, $F(q)\to0$ as $q\to\infty $ by the Riemann--Lebesgue lemma; since the series converges absolutely, every term vanishes and $p(t)\to 1-2/\pi$.
	
	\subsection{Gaussian frequencies}
	For a Gaussian distribution we have $F(q)=\exp({i\overline{\omega} q-q^2\sigma_{\omega}^2/2})$, thus we get
	\begin{equation}
		p(t)=1 - \frac{2}{\pi} + \frac{4}{\pi}\sum_{k=1}^{\infty}\frac{(-1)^k}{4k^2-1}\cos(2\overline{\omega} k t) e^{-2\sigma_{\omega}^2 t^2k^2}.
	\end{equation}
	Note that for large $t$ the terms of the series are rapidly suppressed. At leading order we have
	\begin{equation}
		p(t) \sim 1- \frac{2}{\pi}-\frac{4}{3\pi}\cos(2\overline{\omega} t) e^{-2 \sigma_{\omega}^2 t^2}.
		\label{eq:asymptotic_gaussian}
	\end{equation}
	The parameters of the Gaussian only come into play in the pre-asymptotic regime, in particular in the characteristic time $\tau = 1/\sqrt{2 \sigma_{\omega}^2}$, while the asymptotic value is $p_\infty=1-2/\pi$.
	
	\subsection{Discrete Bernoulli frequencies}
	\label{appendix:average_fraction_bernoulli}
\begin{figure}[t]
		\centering
		\includegraphics[width=0.98\textwidth]{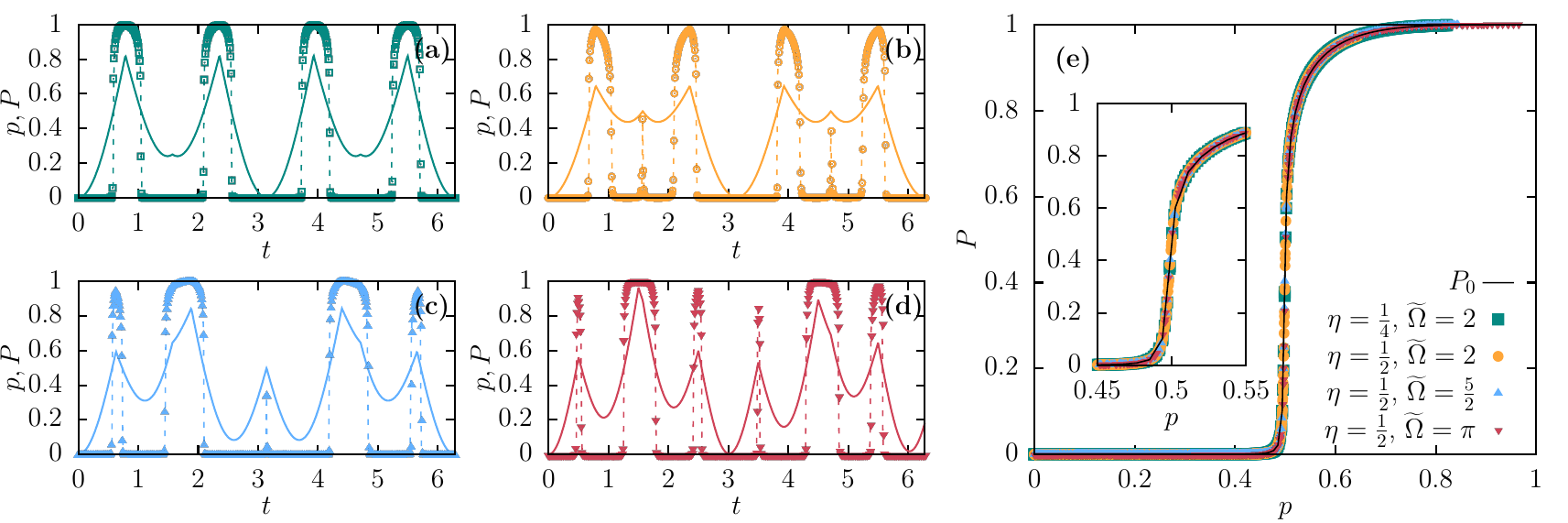}
		\caption{\textbf{Dynamical entanglement percolation observables on a two-dimensional square lattice with i.i.d. Bernoulli frequencies.} 
			(a)-(d) The observables $p(t)$ (solid lines) and $P(t)$ (symbols and dashed lines) for Bernoulli frequencies, with (a) $\eta=1/4$, $\widetilde{\Omega}=2$, (b) $\eta=1/2$, $\widetilde{\Omega}=2$, (c) $\eta=1/2$, $\widetilde{\Omega}=5/2$, (d) $\eta=1/2$, $\widetilde{\Omega}=\pi$. In (a) and (b) the evolution is periodic with period $\pi$; in (c) the evolution is periodic with period $2\pi$; in (d) the evolution is quasi-periodic since $\widetilde{\Omega}$ is irrational. (e) Parametric plot of $P(p)$ using the observables $p(t)$ and $P(t)$ in panels (a)-(d). As for the Gaussian frequencies of Figure~\ref{fig:uncorrelated_noise}, the effect of the dynamics disappears when observing the order parameter $P(p)$. The inset in panel (e) shows the behaviour of $P(p)$ close to the bond percolation threshold $p_c=1/2$. All data perfectly overlap on the order parameter of uniform bond percolation $P_0(p)$ (black line). For all plots, the system size is $L=10^3$, hence $N=10^6$. All results are averaged over $10$ realizations of the disorder, and over $100$ realizations of the edge activation process.}
		\label{fig:uncorrelated_noise_bernoulli}
	\end{figure}

	We assume now that $\omega_{e}$ can take values only in a finite set. The simplest scenario we can imagine is a coin toss: $\omega=\Omega_1$ with probability $\eta$ and $\omega=\Omega_2$ with probability $1-\eta$. In this case we have
	\begin{equation}
		\label{eq:bernoulli}
		\mathcal{P}_{\omega}(x)=\eta \delta(x-\Omega_1)+(1-\eta)\delta(x-\Omega_2),
	\end{equation}
	where $\delta(x)$ is the Dirac delta. Setting $\eta=0$ or $\eta=1$ we recover the uniform case. Frequencies have expected value $\mathbb{E}[\omega]= \eta \Omega_1+(1-\eta)\Omega_2$ and variance $\Var{\omega}=\eta(1-\eta)(\Omega_1-\Omega_2)^2$. The average fraction of maximally entangled states is given by Eq.~\eqref{eq:average_fraction_bernoulli}. The function $|\cos(\Omega_i t)|$ is periodic with period $T_i=\pi/\Omega_i$. Thus $p(t)$ is periodic with period $T$ if and only if there exist two integers $n_1$ and $n_2$ such that $T=n_1 T_1 = n_2 T_2$, which in turn requires $\widetilde{\Omega}=\Omega_2/\Omega_1$ to be a rational number. In particular, if $\widetilde{\Omega} = l/m$, with $l, m$ coprime integers, then the period is $T=m \pi/\Omega_1$. If instead $\widetilde{\Omega}$ is an irrational number, then the function $p(t)$ is quasi-periodic. In both cases, however, the function $P(p)$ is well-defined, indicating that this process is equivalent to uniform bond percolation with probability $p(t)$, as shown in Figure~\ref{fig:uncorrelated_noise_bernoulli}.
	
	\section{Distance distribution in a perturbed lattice}
	\label{appendix:distances}
	
	In order to characterize the statistical properties of the frequencies, we need to determine $\mathcal{P}_d$, as we can use it to compute $\mathcal{P}_{\omega}$. With no loss of generality, let us take the origin at the unperturbed position of $A$ and rotate the $x$-axis along the direction of the unperturbed position of $B$, see for instance nodes $A=0$ and $B=2$ in Figure~\ref{fig:correlations}(a). In other words, the positions of nodes $A$ and $B$ are $\vec{x}_A=(\delta x_{A},\delta y_{A})$ and $\vec{x}_{B}=(1+\delta x_{B},\delta y_{B})$, respectively. The difference $\vec{\Delta}_{AB}=\vec{x}_{B}-\vec{x}_{A}$ is given by $\vec{\Delta}_{AB}=\vec{\nu}+(\Xi_{AB},\Theta_{AB})$, where $\vec{\nu}=(1,0)$ and $\Xi_{AB}=\delta x_{B}-\delta x_{A},\Theta_{AB}=\delta y_{B}-\delta y_{A} \sim \mathcal{N}(0,2\sigma^2)$, since they are differences of two independent Gaussian random variables with zero mean and variance $\sigma^2$. The modulus of this quantity is the distance between $A$ and $B$, $d_{AB}=|\vec{\Delta}_{AB}|$, and it follows a Rice distribution~\cite{rice1944mathematical}
	\begin{equation}
		\label{eq:Rice}
		\mathcal{R}(x;\nu,\sqrt{2}\sigma) = \frac{x}{2\sigma^2} e^{-\frac{x^2+\nu^2}{4\sigma^2}}I_0\left(\frac{x \nu}{2\sigma^2} \right),
	\end{equation}
	where $I_0(x)$ is the modified Bessel function of the first kind with order zero~\cite{abramowitz1965handbook}. Since we are measuring distances in units of lattice spacing, $\nu=1$, we have
	\begin{equation}
		\label{eq:P_d_Rice}
		\mathcal{P}_d(x)=\mathcal{R}(x;1,\sqrt{2}\sigma).
	\end{equation}
	A change of variable gives us $\mathcal{P}_{\omega}$.
	
	\section{Frequency correlations for the threshold model}
	\label{appendix:theta}
	
\begin{figure}[t]
		\centering
		\includegraphics[width=0.95\textwidth]{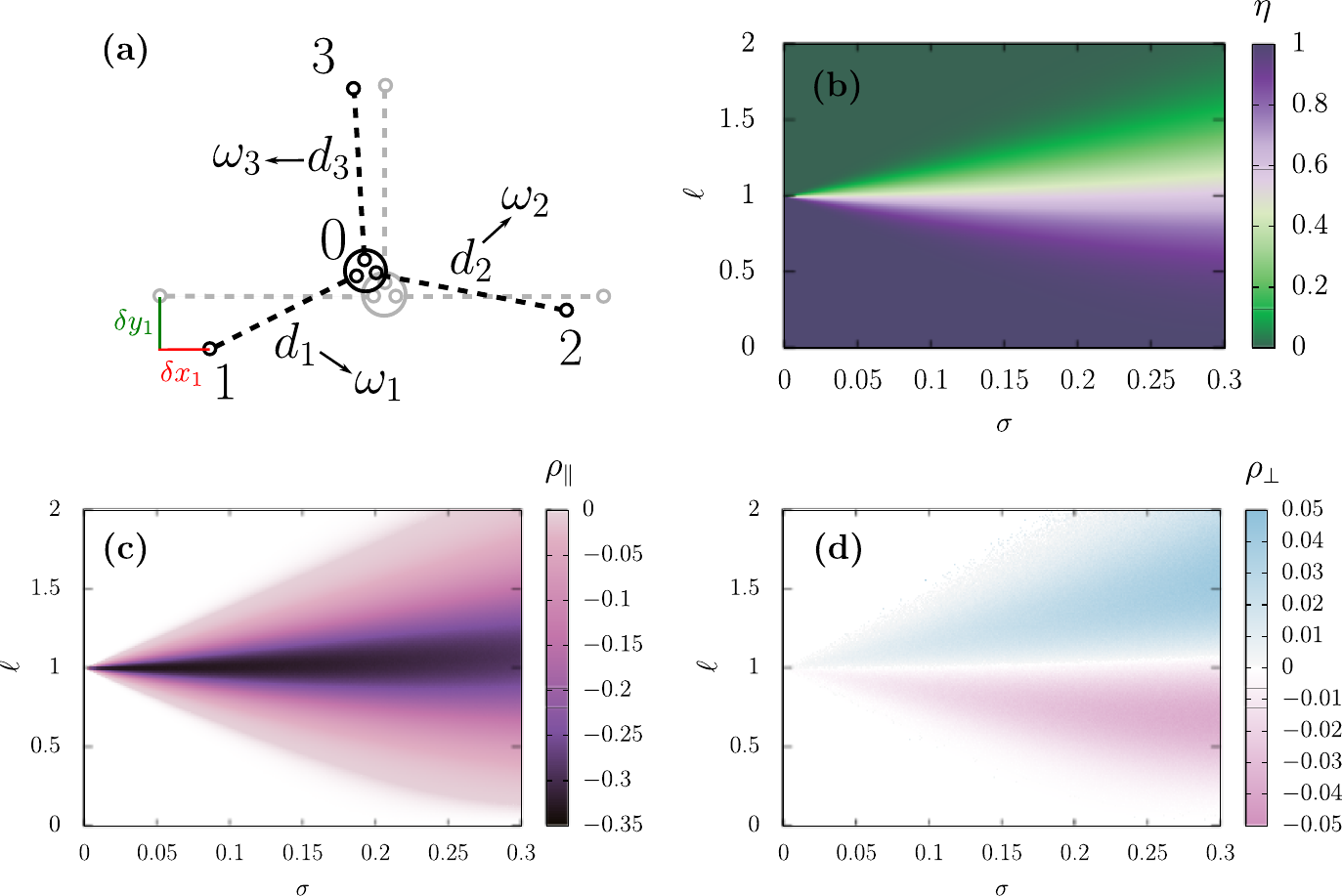}
		\caption{\textbf{Spatial correlations in the perturbed lattice.} (a) Four nodes in the perturbed lattice. Original unperturbed lattice positions are represented with lighter grey lines. For the sake of simplicity, we only represented one qubit on stations $1$, $2$, and $3$. The position of each point is shifted by the independent Gaussian variables $\delta x$ and $\delta y$ (see for instance red and green lines for station $1$). Edges $(1,0)$ and $(0,2)$ are said to be collinear, while $(1,0)$ and $(0,3)$ are perpendicular.
			(b) Numerical evaluation of the probability $\eta(\sigma,\ell)$. The white line corresponds to the median of the distribution, {\sl i.e.}, where $\eta=1/2$. Numerical evaluation of (c) $\rho_{\parallel}$
			and (d) $\rho_{\perp}$ for the two-frequency system, as a function of the noise strength $\sigma$ and the threshold distance $\ell$. Collinear frequencies are anti-correlated, correlations being stronger for $\ell \approx 1$. Correlations between perpendicular frequencies are instead much weaker. Each point is averaged over $10^6$ random realizations of the scenario described in panel (a).}
		\label{fig:correlations}
	\end{figure}

	As already noted in the main text, if frequencies are assigned according to Eq.~\eqref{eq:frequencies_threshold}, {\sl i.e.}, $\omega=1$ for $d>\ell$ and $\omega=\widetilde{\Omega}$ for $d<\ell$, they are Bernoulli variables with parameter $\eta=\mathbb{P}[d > \ell]=\mathbb{P}[\omega=1]$. Figure~\ref{fig:correlations}(b) shows the behaviour of this quantity as a function of $\sigma$ and $\ell$. In particular, for small $\sigma$ the median of $d$ is close to $1$, so that $\eta\approx 1/2$ along the line $\ell \approx 1$, while $\eta$ rapidly tends to $0$ above this line and to $1$ below it, since $d$ is essentially a Gaussian variable fluctuating around $\mathbb{E}[d]$. In this appendix we also discuss the correlation between frequencies of adjacent edges. 
	
	Imagine three nodes aligned in the unperturbed lattice. To simplify the problem, let us also imagine that, instead of Gaussian perturbations, each point gets shifted to the left or to the right by a fixed amount $\delta$. There are eight equally likely scenarios, corresponding to the various left/right shifts. In two of them, when the three nodes are all shifted in the same direction, neither distance is altered. In four of them only one of the two distances changes, while in the remaining two one edge becomes shorter and the other longer. Crucially, in no configuration do both edges become shorter, or both longer: a direct computation gives $\Cov{d_1,d_2}=-\delta^2$. Thus we expect, from this naive argument, that the distances along the same direction should be anti-correlated. We can quantify this intuition, and carry out computations explicitly under a small noise assumption. Let us consider three nodes ($1$, $0$ and $2$) which are three collinear neighbours, and another node ($3$) that is neighbour of node $0$ along the orthogonal direction, see Figure~\ref{fig:correlations}(a). With no loss of generality let us place $0$ in the origin. Using the notation of the previous appendix, $\Xi_{AB}=\delta x_B-\delta x_A$ and $\Theta_{AB}=\delta y_B-\delta y_A$, we then have for the distances
	\begin{align}
		d_{1}&=d_{10}=\sqrt{(1+\Xi_{10})^2+\Theta_{10}^2} \approx 1 +\Xi_{10},\\
		d_{2}&=d_{20}=\sqrt{(1-\Xi_{20})^2+\Theta_{20}^2} \approx 1 - \Xi_{20},\\
		d_{3}&=d_{30}=\sqrt{\Xi_{30}^2+(1-\Theta_{30})^2} \approx 1-\Theta_{30}.
	\end{align}
	Since $\Xi_{10}$ and $\Xi_{20}$ share the variable $\delta x_0$, while $\Xi_{10}$ and $\Theta_{30}$ are independent, it follows that $\Cov{d_{1},d_{2}} \approx -\sigma^2$, and $\Cov{d_{1},d_{3}} \approx 0$, at lowest order in $\sigma$. Thus for low noise we have indeed anti-correlation for the lengths of collinear edges, and no correlations for the lengths of perpendicular edges.

	To quantify the amount of correlations, we introduce the joint probability densities $P_{d_1,d_2|\parallel}(x_1,x_2)$ and $P_{d_1,d_2|\perp}(x_1,x_2)$ for the lengths of two adjacent collinear and perpendicular edges, respectively. From these densities, we can compute the joint distributions of the frequencies $P_{\omega_1,\omega_2| \parallel}(x_1,x_2)$ and $P_{\omega_1,\omega_2|\perp}(x_1,x_2)$, using a simplified notation where $\omega_{i}$ denotes the frequency of the edge of length $d_i$, as in Figure~\ref{fig:correlations}, and the Pearson's coefficients
	\begin{equation}
		\label{eq:pearson_def}
		\rho_{X} = \frac{\Cov{\omega_1,\omega_2| X}}{\Var{\omega }} = \frac{\int \int d x_1 d x_2 P_{\omega_1,\omega_2|X}(x_1,x_2) x_1 x_2 -\mathbb{E}[\omega]^2}{\int dx \mathcal{P}_{\omega}(x) x^2 -\mathbb{E}[\omega]^2} = \frac{\mathbb{E}\left[\omega_1 \omega_2|X \right]-\mathbb{E}[\omega]^2}{\mathbb{E}[\omega^2]-\mathbb{E}[\omega]^2},
	\end{equation}
	with $X=\parallel$ for the collinear case, and $X=\perp$ for the perpendicular case, see Figure~\ref{fig:correlations}(a). Since no closed-form expression is available for these joint densities, we need to resort to numerical methods. However, for the two-frequency case everything turns out to be much simplified, as only two parameters, $\eta$ and $\beta_X$ defined below, are needed to completely characterize correlations for each geometry. Since the value of the frequency $\omega$ depends on whether $d>\ell$ or not, for the two frequencies of adjacent edges -- note that there is no need to distinguish between collinear and perpendicular at this stage -- we can have: (i) both $\omega_1=1$ and $\omega_2=1$, which happens if both $d_1>\ell$ and $d_2>\ell$, thus with probability $\mathbb{P}_{>,>}=\mathbb{P}[d_1>\ell,d_2>\ell]$; (ii) $\omega_1=1$ and $\omega_{2}=\widetilde{\Omega}$, with probability $\mathbb{P}_{>,<}$; (iii) $\omega_1=\widetilde{\Omega}$ and $\omega_{2}=1$, with probability $\mathbb{P}_{<,>}$; (iv) both $\omega_1=\omega_2=\widetilde{\Omega}$, with probability $\mathbb{P}_{<,<}$. In summary, we have
	\begin{align}
		\nonumber
		P_{\omega_1,\omega_2}(x_1,x_2)&=\mathbb{P}_{>,>}\delta(x_1-1)\delta(x_2-1)+\mathbb{P}_{>,<}\delta(x_1-1)\delta(x_2-\widetilde{\Omega})\\
		&+\mathbb{P}_{<,>}\delta(x_1-\widetilde{\Omega})\delta(x_2-1)+\mathbb{P}_{<,<}\delta(x_1-\widetilde{\Omega})\delta(x_2-\widetilde{\Omega}).
	\end{align}
	By normalization, these four probabilities must sum up to $1$. Furthermore, by symmetry $\mathbb{P}_{<,>}=\mathbb{P}_{>,<}$. Finally, note that $\mathbb{P}_{>,>}+\mathbb{P}_{>,<}=\eta$ (equivalently, $\mathbb{P}_{<,<}+\mathbb{P}_{<,>}=1-\eta$), as we have integrated out the variable $d_2$ and we are left with the marginal on $d_1$. Calling $\beta=\mathbb{P}_{<,<}$, we need only the parameters $\eta$ and $\beta$ to characterize the joint distribution and thus the correlations. Note that the geometric constraint of considering collinear rather than perpendicular edges enters in the explicit evaluation of $\beta$. When we need to specify which case we are dealing with, we denote by $\beta_{\parallel}$ and $\beta_{\perp}$ the collinear and perpendicular cases, respectively. Since we have $\mathbb{P}_{<,<}=\beta$, $\mathbb{P}_{<,>}=\mathbb{P}_{>,<}=1-\eta-\beta$, and $\mathbb{P}_{>,>}=\beta+\eta-(1-\eta)$, we can compute
	\begin{equation}
		\mathbb{E}[\omega_1 \omega_2] = \mathbb{P}_{>,>} + 2 \widetilde{\Omega} \mathbb{P}_{<,>} + \widetilde{\Omega}^2 \mathbb{P}_{<,<} 
		= \beta (\widetilde{\Omega}-1)^2 + \eta - (1-\eta)+2(1-\eta)\widetilde{\Omega},
	\end{equation}
	and, subtracting $\mathbb{E}[\omega]^2=[\eta+(1-\eta)\widetilde{\Omega}]^2$, we get
	\begin{equation}
		\Cov{\omega_1,\omega_2} = (\widetilde{\Omega}-1)^2[\beta -(1-\eta)^2].
	\end{equation}
	Using $\Var{\omega}=\eta(1-\eta)(\widetilde{\Omega}-1)^2$, if $\widetilde{\Omega}\neq 1$ -- otherwise $\Cov{\omega_1,\omega_2}=\Var{\omega}=0$ -- we obtain from Eq.~\eqref{eq:pearson_def}
	\begin{equation}
		\rho_{X} = \frac{\beta_{X}-(1-\eta)^2}{\eta(1-\eta)}.
	\end{equation}
	Note that this expression does not depend on $\widetilde{\Omega}$. It is just a measure of the tendency of two adjacent frequencies to be similar. If there are no spatial constraints, we have $\beta_{X}=(1-\eta)^2$ and $\rho_{X}=0$. However, as we have already noted, we expect collinear distances to be negatively correlated. Since at lowest order in $\sigma$ the lengths are jointly Gaussian, this implies that the probability that both are shorter than $\ell$ is smaller than for independent lengths, {\sl i.e.}, $\beta_{\parallel}<(1-\eta)^2$. Thus we expect $\rho_{\parallel} < 0$. On the other hand, perpendicular lengths are uncorrelated at lowest order in $\sigma$, so only weak correlations are expected for perpendicular edges. This is confirmed by means of numerical evaluation of $\rho_{X}$, as reported in Figure~\ref{fig:correlations}(c)-(d), obtained averaging over $10^6$ random realizations of the four-node scenario described in Figure~\ref{fig:correlations}(a). Alternatively, one can compute $\eta$ using the Gaussian small-noise approximation of the Rice distribution~\cite{abramowitz1965handbook}, and $\beta$ by Monte Carlo integration.
	
	\section{Mean-field solution for the two-colour bond percolation}
	\label{appendix:mean_field}
	
	The basic idea to derive a mean-field solution for a percolation process is to
	consider the process on a random network, ignore the presence of short loops
	and use the formalism of branching processes~\cite{dorogovtsev2022nature}.
	To model the case of square lattice, we consider a random regular graph with coordination number $k=4$, and each node has attached exactly two edges of each type. This graph model is a particular case of a more general coloured configuration model~\cite{kryven2019bond}.
	Let us denote by $m_1$ the probability that if we follow an active edge of type $1$ we reach the giant component (GC). $m_2$ is defined analogously. An active edge leads to the GC if at least one of the outgoing edges leads to the GC. Ignoring the presence of loops we can assume that edge probabilities are independent -- and here lies the tree-like approximation. Thus we can write
	\begin{align}
		\label{eq:m1}
		m_1 &= 1 - (1-\phi_1 m_1)(1-\phi_2 m_2)^2,\\
		\label{eq:m2}
		m_2 &= 1- (1-\phi_1 m_1)^2(1-\phi_2 m_2).
	\end{align}
	The stable fixed point of these equations $(m_{1}^*,m_{2}^*)$ gives us the probability that following an active edge we reach the GC.
	A similar argument allows us to get an equation for the size of the GC. A node is in the GC if at least one of its edges leads to it, and this occurs with probability
	\begin{equation}
		S = 1-(1-\phi_{1} m_{1}^{*})^2(1-\phi_{2} m_{2}^{*})^2.
	\end{equation}
	Thus the stable fixed point of Eqs.~\eqref{eq:m1} and~\eqref{eq:m2} plugged into the expression for $S$ gives us the order parameter $S(\phi_1,\phi_2)$.

	\subsection{Percolation threshold}
\begin{figure}
		\centering
		\includegraphics[width=0.95\textwidth]{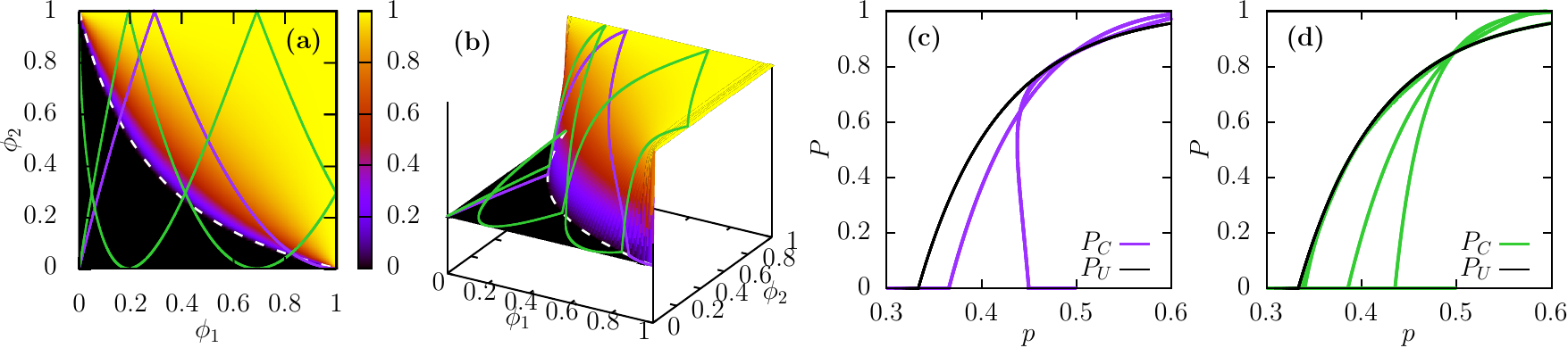}
		\caption{\textbf{Mean-field phase diagram for the two-colour bond percolation.} (a) The phase diagram in the parameter plane $\phi_1-\phi_2$. The dashed line corresponds to the critical threshold Eq.~\eqref{eq:mean_field_critical}, the continuous lines are two dynamical evolutions as in Eq.~\eqref{eq:dynamical_activation} with $\widetilde{\Omega}=2$ (purple) and $\widetilde{\Omega}=5/2$ (green). Arrows indicate the direction of time up to half of the period. (b) 3D plot of the surface $S(\phi_1,\phi_2)$. Lines are the evolution of $S$ along the curves in panel (a). (c) Parametric plot for the order parameter $P_C(p)$ (symbols) and $P_U(p)$ (black line) for $\widetilde{\Omega}=2$ (purple). (d) Same as in (c) but for $\widetilde{\Omega}=5/2$. The overall picture is qualitatively analogous to what we observe in numerical simulations on two-dimensional lattices, Figure~\ref{fig:two_colour_phase}.}
		\label{fig:mean_field}
	\end{figure}

	Note that $m^{*}_1=m^{*}_2=0$ is always a trivial solution of the fixed-point equations, and it corresponds to $S=0$, {\sl i.e.}, no GC exists. However, for $\phi_1=\phi_2=1$ we have another solution $m_1=m_2=1$. Thus there must exist a critical value of the parameters $\phi_1,\phi_2$ separating a phase with no GC from a phase with a GC. This defines a percolation line. We can determine such a critical line studying the stability of the trivial fixed point to determine where another stable solution with positive probabilities appear. The largest eigenvalue of the Jacobian matrix, of Eqs.~\eqref{eq:m1}-\eqref{eq:m2}, evaluated at $(0,0)$, is given by
	\begin{equation}
		\Lambda(\phi_{1},\phi_{2}) =  \frac{\phi_{1}+\phi_{2} + \sqrt{\phi_{1}^2 + \phi_{2}^2 +14\phi_{1} \phi_{2}}}{2}.
	\end{equation}
	The trivial fixed point $(0,0)$ remains stable until $\Lambda=1$. This condition defines a critical line in the $\phi_{1} - \phi_{2} $ plane determined by the equation
	\begin{equation}
		3 \phi_1 \phi_{2} + \phi_{1} + \phi_{2} - 1 = 0,
		\label{eq:mean_field_critical}
	\end{equation}
	see the dashed line in Figure~\ref{fig:mean_field}(a) which separates a phase with no GC (below) from a phase with a GC (above).
	
	\subsection{Dynamical evolution for the activation probabilities}
	As shown in the main text, the case of activation probabilities that evolve in time is described by a curve $(\phi_1(t),\phi_2(t))$ in the parameter space, see the continuous lines in Figure~\ref{fig:mean_field}(a). Thus, the order parameter $P(t)$ is obtained as $P(t)=S(\phi_1(t),\phi_2(t))$. Clearly, the local constraint in the problem results in an order parameter that cannot be expressed in terms of the fraction of active edges $p$.
	
	\subsection{Random reshuffling}
	Using the analytical mean-field solution, we can analyse what happens if we perform a random reshuffling of the edge colours, destroying the correlation structure. In such a case, we have a global constraints fixing half of the edges of each type. Thus there is no local constraint on the edges attached to each node. In other words, each edge is either of type 1 or 2 with probability $1/2$, independently of its neighbours. Thus Eq.~\eqref{eq:m1} and Eq.~\eqref{eq:m2} denote the probability of one particular configuration of the possible realizations of a node's neighbourhood. Averaging over all possible configurations we get
	\begin{equation}
		m_1 = 1-\frac{1}{2^3}\sum_{n=0}^{3}{3 \choose n} (1-\phi_1 m_1)^{n}(1-\phi_2 m_2)^{3-n} = 1-\left(1- \frac{\phi_1 m_1 + \phi_2 m_2}{2} \right)^3,
	\end{equation}
	and correspondingly for $m_2$. Since it follows that $m_1=m_2=m$, the branching process can be solved using just one recursive equation
	\begin{equation}
		m = 1-(1-pm)^3,
	\end{equation}
	where $p=(\phi_1+\phi_2)/2$ is the average fraction of active edges. The same argument leads to an equation for the order parameter
	\begin{equation}
		P_{U}=1-(1-p m)^4.
	\end{equation}
	These two last equations are the well-known solution for uniform bond percolation on a random regular graph with activation probability $p$. This shows explicitly, in agreement with Ref.~\cite{romancino2026entanglement}, that the role of the microscopic disorder disappears as soon as edge correlations are removed.
	
	\twocolumngrid
	
	\end{document}